# Ultrafast Optical Spectroscopy Evidence of Pseudogap and Electron-Phonon Coupling in an Iron-Based Superconductor KCa$_2$Fe$_4$As$_4$F$_2$

Chen Zhang,[1, *] Qi-Yi Wu,[1, *] Wen-Shan Hong,[2, 3, *] Hao Liu,[1] Shuang-Xing Zhu,[1] Jiao-Jiao Song,[1] Yin-Zou Zhao,[1] Fan-Ying Wu,[1] Zi-Teng Liu,[1] Shu-Yu Liu,[1] Ya-Hua Yuan,[1] Han Huang,[1] Jun He,[1] Shiliang Li,[2,4] H. Y. Liu,[5] Yu-Xia Duan,[1] Hui-Qian Luo,[2,4, †] and Jian-Qiao Meng[1, ‡]

[1]*School of Physics and Electronics, Central South University, Changsha 410083, Hunan, China*
[2]*Beijing National Laboratory for Condensed Matter Physics,
Institute of Physics, Chinese Academy of Sciences, Beijing 100190, China*
[3]*School of Physical Sciences, University of Chinese Academy of Sciences, Beijing 100049, China*
[4]*Songshan Lake Materials Laboratory, Dongguan 523808, China*
[5]*Beijing Academy of Quantum Information Sciences, Beijing 100193, China*
(Dated: Friday 21$^{st}$ January, 2022)

We use ultrafast optical spectroscopy to study the nonequilibrium quasiparticle relaxation dynamics of the iron-based superconductor KCa$_2$Fe$_4$As$_4$F$_2$ with $T_c$ = 33.5 K. Our results reveal a possible pseudogap ($\Delta_{PG}$ = 2.4 ± 0.1 meV) below $T^* \approx$ 50 K but prior to the opening of a superconducting gap ($\Delta_{SC}(0) \approx$ 4.3 ± 0.1 meV). Measurements under high pump fluence reveal two distinct, coherent phonon oscillations with 1.95 and 5.51 THz frequencies, respectively. The high-frequency $A_{1g}(2)$ mode corresponds to the $c-$axis polarized vibrations of FeAs planes with a nominal electron-phonon coupling constant $\lambda_{A_{1g}(2)}$ = 0.194 ± 0.02. Our findings suggest that the pseudogap is likely a precursor of superconductivity, and the electron-phonon coupling may play an essential role in the superconducting pairing in KCa$_2$Fe$_4$As$_4$F$_2$.

**KCa$_2$Fe$_4$As$_4$F$_2$, ultrafast spectroscopy, electron-phonon coupling, superconductivity**

## 1 INTRODUCTION

The microscopic mechanism of high-temperature superconductivity (HTSC) has been debated for decades. Due to the intertwined interactions of charge, orbital and spin freedoms, it is a great challenge to elucidate the key issue in such a strongly correlated electron system [1, 2]. Although the carriers for superfluid in HTSC are still regarded as the electron pairs (Cooper pair) as first proposed in the conventional BCS superconductors based on electron-phonon ($e$-$ph$) coupling [3], both the pairing interaction and condensation process remain unanswered [4, 5]. In copper oxides, spin fluctuation is the possible pairing glue as supported by the universal neutron spin resonance mode among all systems [6, 7], the $e$-$ph$ coupling may trigger Cooper pair formation [8], but the $e$-$ph$ coupling constant, $\lambda$, is too weak to maintain such high $T_c$ superconductivity [9–11]. Surprisingly, it appears that the Cooper pairs can form above $T_c$, resulting in a pseudogap phase, either in competition with or as a precursor to the superconducting gap below $T_c$ [8]. Such questions still remain to resolve in iron-based superconductors (FeSC), where similar competing orders and collective spin resonance modes have been observed in numerous family members [12–15]. However, the $e$-$ph$ coupling has never been ruled out in electron pairing of FeSC; for instance, it possibly boosts $T_c$ in FeSe monolayer thin film [16]. So far, there are some experimental traces but no solid evidence for the pseudogap phase in FeSC, as it may have a much smaller magnitude compared with cuprates [17–19].

Ultrafast optical pump-probe spectroscopy is a an effective method for studying quasiparticle (QP) dynamics and HTSC $e$-$ph$ couplings [20–24]. Different QP components in post-laser pulse excitations can be distinguished in the temporal domain based on their varying lifetimes. Since the energy transferred from nonequilibrium QP to the lattice *via* the $e$-$ph$ scattering channel is the primary electron relaxation process, the $e$-$ph$ coupling constant $\lambda$ is extractable from $e$-$ph$ scattering times. In iron-based superconductors, the phonon vibration mode, particularly the $A_{1g}$ mode, is more readily observed using ultrafast spectroscopy than that in cuprates with strongly correlated states. Due to the slower relaxation process caused by the bottleneck effect, ultrafast spectroscopy is more sensitive to measuring small gaps than normal angle-resolved photoemission spectroscopy (ARPES). We can readily estimate the gap size by following the temperature dependence of amplitudes and relaxation time associated with superconductivity using the Rothwarf-Taylor (RT) model [24–28].

Here in this letter, we report an ultrafast optical spectroscopy study on KCa$_2$Fe$_4$As$_4$F$_2$ (12442), a quasi-two-dimensional iron-based superconductor with $T_c$ =33.5 K and asymmetric Fe$_2$As$_2$ bilayer structure similar to cuprates [29–31]. The 12442 compound is a stoichiometric and self-hole doped system with no magnetic or structural transitions, making it a suitable platform for studying energy gaps with preserved symmetry both on spin and lattice structure [30, 32]. The gap function has been investigated by a variety of techniques, including $\mu$SR [33], specific heat [34], nuclear magnetic resonance(NMR) [35], optical spectroscopy [36], ARPES [37] and scanning tunneling microscopy/spectroscopy (STM/STS) [38, 39] measurements. ARPES and STM/STS measurements suggest a great diversity of superconducting gaps. Transport [40], NMR [35], and optical [41] experiments suggest a pseudogap opening above $T_c$. In our high-resolution ultrafast spectroscopy ex-periments, a pseudogaplike feature ($\Delta_{PG} = 2.4 \pm 0.1$ meV) is detected from $T_c$ to $T^* \approx 50$ K; its amplitude reduces substantially below $T_c$ due to the opening of the superconducting gap ($\Delta_{SC}(0) \approx 4.3 \pm 0.1$ meV). We also identity two separate coherent phonon oscillations [$A_{1g}(1)$ and $A_{1g}(2)$] and calculate the electron-phonon coupling constant $\lambda_{A_{1g}(2)} = 0.194 \pm 0.02$ for the higher frequency $A_{1g}(2)$ mode. Our results suggest that both the precursory pseudogap and phonon mode may play essential roles in the superconducting pairing in KCa$_2$Fe$_4$As$_4$F$_2$.

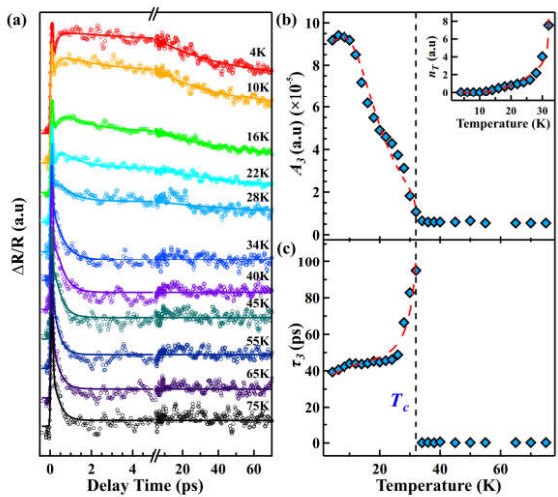

FIG. 1. (color online) **(a)** $\Delta R(t)/R$ as a function of delay time over a temperature range from 4 to 75 K at pump fluence $\sim 4.2$ μJ/cm². The solid lines are Eq. (1) fits. **(b)** $A_3$ temperature dependence. **(c)** $\tau_3$ temperature dependence. The inset (b) shows thermally-excited quasiparticle, $n_T(T)$, densities. The dashed red lines show the RT model fitting curves.

## 2 EXPERIMENTAL DETAILS

In this study, high-quality single crystal KCa$_2$Fe$_4$As$_4$F$_2$ was grown using the self-flux method [15]. Ultrafast time-resolved differential reflectivity $\Delta R/R$ was measured at a center wavelength of 800 nm ($\sim 1.55$ eV) using a 1 MHz Yb-based femtosecond (fs) laser oscillator (Supplemental Material) [42, 43]. Measurements were performed on a freshly-cleaved surface under a $10^{-6}$ mbar vacuum.

## 3 RESULTS AND DISCUSSION

Typical reflectivity signal, $\Delta R/R$, was a function of delay time over a temperature range from 4 to 75 K [Fig. 1(a)]. The temperature-dependent transient reflectivity was measured at a modest fluence of $\sim 4.2$ μJ/cm². Transient reflectivity is dominated by electron-electron (e-e) and electron-boson scattering processes; thus, the relaxation processes can be well fitted with three-exponential (Supplemental Materials) decays convoluted with a Gaussian laser pulse:

$$\frac{R(t)}{R} = \frac{1}{\sqrt{2\pi}w}\exp(-\frac{t^2}{2w^2}) \otimes [\sum_{i=1}^{3} A_i\exp(-\frac{t-t_0}{\tau_i})] + C \quad (1)$$

where $A_i$ is the amplitude, $\tau_i$ is the relaxation time of the $i$th decay process, $w$ is the incidence pulse temporal duration, and $C$ is a constant offset. The first and briefest relaxation process, whose lifetime $\tau_1$ is comparable to the instrument temporal resolution, is generally considered as an e-e scattering process [44].

Figs. 1(b) and (c) summarize the temperature dependence of $A_3$ and $\tau_3$ of the third component. Both show an abrupt change near $T_c$. $A_3$ increases as temperature decreases below $T_c$ and gradually saturates at low temperatures, and $\tau_3$ decreases significantly when temperature slips below $T_c$. These features are commonly observed in HTSC materials and can be well described by the RT model [24, 26, 27]. If photon excitation triggers a system with a superconducting gap ($\Delta_0$), photo-excited carriers relax to the ground state by emitting high-energy phonons with an energy greater than $2\Delta_0$. High-energy phonons reexcite QPs to an unoccupied state. These two processes happen simultaneously and form a dynamic balance. The quasi-equilibrium state retards QP relaxation and is called the bottleneck effect. The balance is only interrupted when high-energy phonons become low-energy phonons with an energy lower than $2\Delta_0$, either through harmonic decay or diffusion into the nonirradiated areas of the material. In a strong bottleneck, QP relaxation times slow down greatly and manifest a sudden change near $T_c$. [Fig. 1(c)]. Thermally-excited QPs, $n_T$, can be obtained via $n_T(T) \sim A(0)/A(T) - 1$ [26, 28]. It greatly increases as temperatures approach $T_c$ [inset of Fig. 1(b)]. The $T$-dependent energy gap $\Delta(T)$ can be extracted from the following formulas,

$$A(T) \propto \frac{\varepsilon_I/[\Delta(T) + k_B T/2]}{1 + \gamma\sqrt{2k_B T/\pi\Delta(T)}e^{-\Delta(T)/k_B T}}, \quad (2)$$

$$\tau_3 \propto \frac{\hbar\omega^2 \ln\{1/[\varepsilon_I/\alpha\Delta(0)^2 + e^{-\Delta(T)/k_B T}]\}}{12\Gamma_\omega\Delta(T)^2}. \quad (3)$$

Here $\varepsilon_I$ is the absorbed laser energy density per unit cell. $\gamma$, $\omega$, $\alpha$, and $\Gamma_\omega$ are the fitting parameters. The fitting results are shown by the dashed red lines in Figs. 1(b) and (c), giving a zero-temperature gap of $\Delta(0) \approx 4.3 \pm 0.1$ meV. The superconducting gap obtained in our study is consistent with recent STM/STS ($4.6 \pm 0.2$ meV) [38, 39] and specific heat results (5.3 meV) [34], as well as the averaged gap value from ARPES [37]. It is worth noting that multiple superconducting gaps are observed in 12442 by STM, specific heat, and ARPES, and the value of $\Delta(0)$ obtained in our study is an average value of multiple superconducting gaps.

Fig. 2(a) is a 2D pseudocolor $\Delta R(t)/R$ mapping image with temperature versus delay time. Besides the superconductivity induced change of $\tau_3$ at $T_c$, the second relaxation process $\tau_2$ also changes greatly below $T \approx 50$ K. Fluence-dependent measurements were performed at various temperatures to gain more component information. Figs. 2(b) and 2(c) show normalized curves at 35 and 50 K, respectively. At 35 K, which is slightly above $T_c$, a strong fluence-dependent relaxation process, $\tau_2$, is still observed during the brief timescale $0.3 \le t \le 2$ ps. Such behavior has gone at 50 K. To quantitatively study $\tau_2$-related QPs relaxation, fluence-dependent data were also fitted with Eq. (1). Fig. 2(d) summarizes decay rate $\gamma_2$ (= $1/\tau_2$) as a function of temperature for different pump fluences; almost all cases suggest a clear fluence-dependence behavior below 50 K. At low fluences, as temperature increases, $\gamma_2$ increases steadily below $T^*$ and keeps constant above $T^*$. Similar fluence-dependent behaviors have been reported on moderately hole-doped $Ba_{1-x}K_xFe_2As_2$ and ascribed to the appearance of pseudogap [45, 46]. Since 12442 shares similar characters, it is reasonable to argue that here in 12442, the pseudogap opening temperature of $T^* \approx 50$ K, and $\tau_2$ is closely related to the pseudogap. Figs. 2(e) and (f) display the temperature dependence of $\tau_2$, $A_2$ and $n_T$ at a pump fluence of $\sim 4.07$ $\mu J/cm^2$. Below $T^*$, $\tau_2$ increases steadily as temperature decreases. $A_2$ first increases between $T^*$ and $T_c$ and then decreases significantly in the SC state, indicating a suppression of the pseudogap signal below $T_c$. According to the methodology of Liu *et al.*'s analysis of the hidden-order gap in $URu_2Si_2$ [47], the RT model can extract the size of the pseudogap of 12442. Therefore, we use Eq. (3) to fit the temperature dependence of $\tau_2$, giving $\Delta_{PG} \approx 2.4 \pm 0.1$ meV. In addition, a pseudogap with a similar size was obtained by fitting the temperature dependence of $n_T$ above $T_c$. Here, the superconducting gap and the pseudogap we obtained are of different sizes. The ultrahigh-resolution ARPES observed multiple Fermi surfaces with different gap sizes [37]. Given that the size of the pseudogap is similar to that of the minimal superconducting gap on $\gamma$ sheets, it is reasonable to consider that the observed small pseudogaplike feature might open in the $\gamma$ band [37], or above the Fermi level. A similar gap size also means that the pseudogap may be the precursor of the superconducting gap.

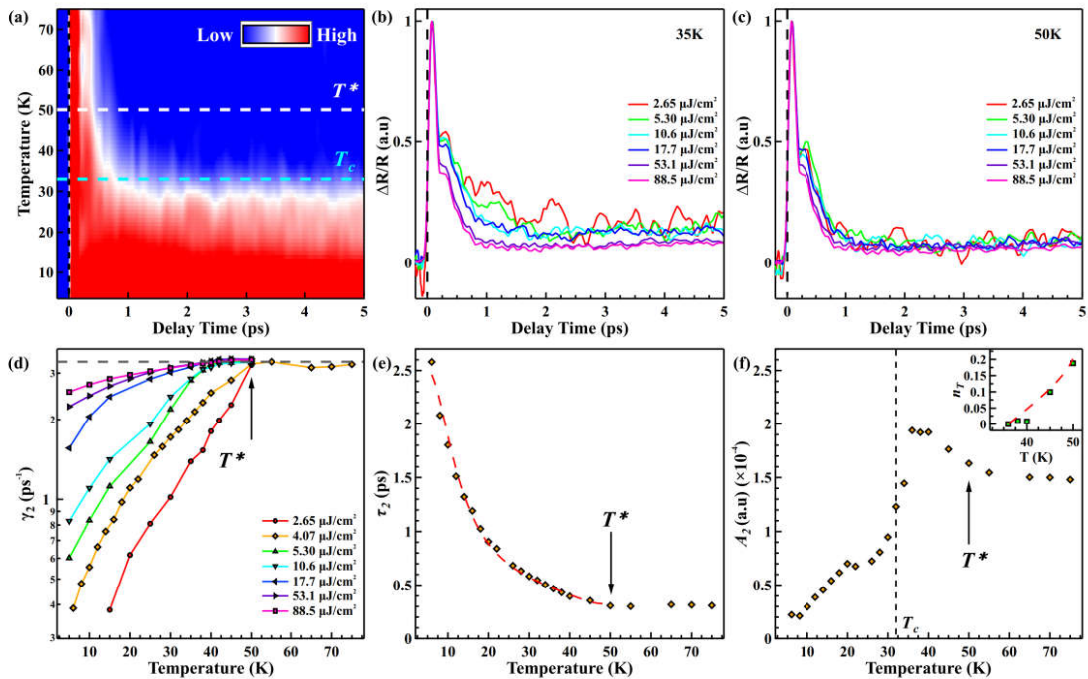

FIG. 2. (color online) **(a)** 2D pseudocolor map of $\Delta R(t)/R$ as a function of temperature and delay time. Black and cyan dash lines indicate $T_c$ and $T^*$, respectively. **(b)** normalized $\Delta R(t)/R$ at 35 K, which is slightly above $T_c$. **(c)** normalized $\Delta R(t)/R$ at 50 K as a function of pump fluence. **(d)** Decay rate, $\gamma_2$, is shown as a function of temperature measured at different fluences. An anomaly at $T^*$ is clear. **(e)** temperature dependence of relaxation time $\tau_2$. **(f)** shows the temperature dependence of the amplitude $A_2$. Inset: the density of thermally-excited QPs $n_T$. The dashed red lines in (e) and (f) are the RT model fit.

Having investigated the behavior of possible pseudogap, we now discuss the *e-ph* coupling constant investigated using temperature-dependent measurements under a high pump fluence of $\sim 84$ $\mu J/cm^2$. Apparent oscillations appear in $\Delta R/R$ mapping [Fig. 3(a)]. The oscillation periods increases as temperature rises. Fig. 3(b) shows a typical $\Delta R/R$ signal at 4 K. It is composed of both coherent oscillations and exponential relaxation fitted by Eq. (1) (red line). Oscillatory components were extracted by performing Fast Fourier transform (FFT) on the oscillation after subtracting nonoscillatory background [42, 43]. The Fig. 3(b) inset presents two pronounced terahertz modes, $\omega_1$ and $\omega_2$, at frequencies 1.95 (i.e., 8.1 meV or 65 $cm^{-1}$) and 5.51 THz (i.e., 22.8 meV or 183.8 $cm^{-1}$). Both modes are $A_{1g}$ modes, labeled $A_{1g}(1)$ and $A_{1g}(2)$, respectively, corresponding to *c*-axis polarized vibrations of FeAs planes [48–50].

The *e-ph* coupling constant, $\lambda$, is usually extracted using a two-temperature model (TTM) based on the assumption that *e-e* scattering time $\tau_{e-e}$ is much shorter than *e-ph* scattering time $\tau_{e-ph}$ [51]. At relatively low temperatures and low laser excitation densities, excited electrons are unable to reach thermal equilibrium by the $e$-$e$ scattering process on the time scale of the $\tau_{e-ph}$. This results in a TTM underestimation of $\lambda$ [52–55]. To describe the situation, an extended multitemperature model (EMTM) has been proposed. This includes $\tau_{e-e}$ which is comparable to $\tau_{e-ph}$ [20, 56]. This model can describe well the $e$-$ph$ scattering of cuprates and iron-pnictides at high temperatures [57–60], where a relationship between the second moment of the Eliashberg spectral function $\lambda \langle \omega^2 \rangle$ and $\tau_{e-ph}$ is :

$$\lambda \langle \omega^2 \rangle = \frac{2\pi k_B T_l}{3\hbar \tau_{e-ph}}. \tag{4}$$

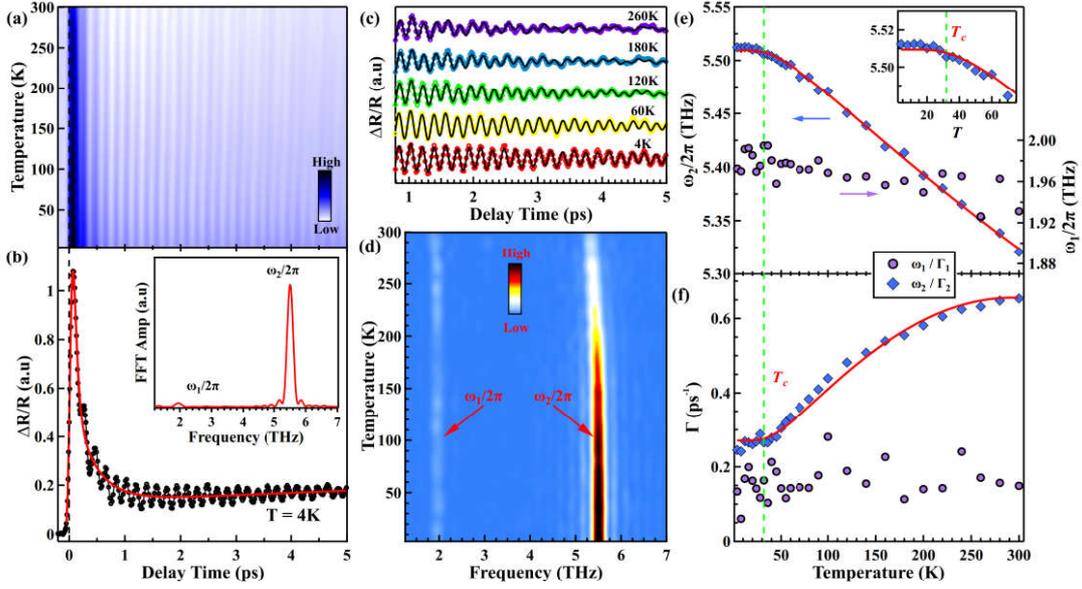

FIG. 3. (color online) (**a**) 2D pseudocolor map of $\Delta R(t)/R$ as a function of temperature and delay time at a pump fluence of $\sim 84$ $\mu J/cm^2$. (**b**) High fluence $\Delta R(t)/R$ at 4 K, showing coherent optical phonon vibration superimposed on QPs relaxation. The solid red line is an exponential fitting result of the QPs relaxation process. Inset: FFT results showing two clear peaks at 1.95 and 5.51 THz. (**c**) phonon oscillations extracted at five different temperatures. Fitted results are in black. (**d**) FFT spectrum color intensity map as a function of frequency and temperature. (**e**) and (**f**) the derived frequency and damping rate, respectively, as a function of temperature using Eq. (5). The solid red lines in (e) and (f) are fitted curves using an anharmonic phonon model. Insst in (e): magnified views.

Here $\omega$ is the phonon frequency, and $T_l$ is the lattice temperature. Usually, $T_l$ is close to the ambient temperature due to the high heat capacity of the lattice. According to Eq. (4), the $\tau_{e-ph}$ is proportional to the sample temperature. The fitting result of the temperature-dependent $\tau_{e-ph}$ follows this linear relationship well above the Debye temperature $\theta_D \sim 240$ K (Supplemental Materials). $\lambda \langle \omega^2 \rangle$ was estimated to be $2.2 \times 10^{26}$ Hz$^2$ (i.e., 95 meV$^2$). Due to the lack of information about phonon density of states, the $\langle \omega^2 \rangle$ value cannot be obtained. Here, the $e$-$ph$ coupling constant is estimated to be $\lambda_{A_{1g}(2)} = 0.194 \pm 0.02$ according to the value of $A_{1g}(2)$ phonon energy $\sim 22.8$ meV. Therefore, the $\lambda_{A_{1g}}$ and $T_c$ of the 12442 system are consistent with the suggested positive correlation between $\lambda_{A_{1g}}$ and $T_c$ in other FeSCs [23].

Fig. 3(c) presents the time-domain oscillations at several temperatures. Oscillations exist at all measured temperatures ranging from 4 to 300 K. Fig. 3(d) displays the FFT spectrum as a function of frequency and temperature. Two distinct terahertz modes, $\omega_1$ and $\omega_2$, were observed at all measured temperatures. Oscillations at room temperature are generally considered to be caused by coherent phonons. $\omega_2$ frequency varies significantly with temperature. $\omega_1$ frequency variances are petty, and its FFT amplitude is much weaker than that of $\omega_2$. Oscillation damping rates and amplitudes were extracted using fitting damped oscillations via the expression [black curves in Fig. 3(c)]

$$\left(\frac{\Delta R}{R}\right)_{osc} = \sum_{j=1,2} A_j e^{-\Gamma_j t} \sin(\omega_j t + \phi_j), \tag{5}$$

where $A_j$, $\Gamma_j$, $\omega_j$, and $\phi_j$ are the $j$th oscillatory signal amplitude, damping rate, frequency, and initial phase, respectively. The extracted temperature evolutions of frequencies and damping rate are plotted in Figs. 3(e) and (f), respectively. $\omega_1$ and $\Gamma_1$ vary little with temperature. $\omega_2$ and $\Gamma_2$ do vary significantly with temperature. $\omega_2$ softens (redshift) as temperature increases, from $\sim 5.51$ THz at 4 K to $\sim 5.34$ THz at 300 K [Fig. 3(e)]. $\omega_2$ and $\Gamma_2$ temperature dependence in the experimental temperature range can be fitted by optical phonon anharmonic effects [61, 62]. This shows that there is no structural phase transition in the temperature range of 4 - 300 K, which is consistent with earlier experimental results [30, 32].

# 4 CONCLUSIONS

In conclusion, ultrafast optical spectroscopy was used to analyze $KCa_2Fe_4As_4F_2$ QPs dynamics. Temperature dependence of QPs relaxation revealed two characteristic temperatures, $T_c$ associated with a superconducting gap opening; and $T^* \approx 50$ K caused by a possible small pseudogap onset. Fitting results gave $\Delta_{SC}(0) \approx 4.3 \pm 0.1$ meV and $\Delta_{PG} = 2.4 \pm 0.1$ meV, respectively. Two coherent phonons were observed at 1.95 and 5.51 THz ($T = 4$ K). The high-frequency $A_{1g}(2)$ mode with $\lambda_{A_{1g}(2)} = 0.194 \pm 0.02$, can be explained by optical phonon anharmonic effects. Our experimental findings shed light on the probable significance of pseudogap and phonon modes in the electron pairing process in iron-based superconductors.

This work was supported by the National Natural Science Foundation of China (Grants No. 51502351, No. 12074436, and No. U2032204, No. 11822411, No. 11961160699, and No. 11874401), the National Key Research and Development Program of China (Grants No. 2018YFA0704200, No. 2017YFA0303100, and No. 2017YFA0302900), the Strategic Priority Research Program (B) of the Chinese Academy of Sciences (CAS) (Grants No. XDB25000000 and No. XDB07020300) and K. C. Wong Education Foundation (GJTD-2020-01). J. Q. M. would like to acknowledge support from the Innovation-driven Plan in Central South University (2016CXS032). H. L. is grateful for the support from the Youth Innovation Promotion Association of CAS (Grant No. Y202001) and the Beijing Natural Science Foundation (Grant No. JQ19002).

---

\* These three authors contributed equally
† Corresponding author: hqluo@iphy.ac.cn
‡ Corresponding author: jqmeng@csu.edu.cn